# Geographical Modeling: from Characteristic Scale to Scaling

Yanguang Chen

(Department of Geography, College of Urban and Environmental Sciences, Peking University, Beijing 100871, P.R.China. E-mail: chenyg@pku.edu.cn)

**Abstract**: Geographical research was successfully quantified through the quantitative revolution of geography. However, the succeeding theorization of geography encountered insurmountable difficulties. The largest obstacle of geography's theorization lies in scale-free distributions of geographical phenomena which exist everywhere. The first paradigm of scientific research is mathematical theory. The key of a quantitative measurement and mathematical modeling is to find a valid characteristic scale. Unfortunately, for many geographical systems, there is no characteristic scale. In this case, the method of scaling should be employed to make a spatial measurement and carry out mathematical modeling. The basic idea of scaling is to find a power exponent using the double logarithmic linear relation between a variable scale and the corresponding measurement results. The exponent is a characteristic parameter which follows a scaleful distribution and can be used to characterize the scale-free phenomena. The importance of the scaling analysis in geography is becoming more and more evident for scientists.

**Key words**: Geographical modeling; Spatial analysis; Characteristic scale; Scaling; Scale-free distribution

# 1 Introduction

The development level of a subject depends heavily on the methods which are applied to this subject. The scientific method include two basic paradigms: one is the *mathematization of the world picture*, and the other, *experience and experiment* (Henry, 2002). Sixty year ago, Einstain (1953) once pointed out: "The development of Western science has been based on two great achievements, the invention of the formal logical system (in Euclidean geometry) by the Greek philosophers，and the discovery of the possibility of finding out causal relationships by systematic experiment (at the Renaissance)." (see Crombie, 1963, page 142) Formal logical system includes mathematics and is related to the relationships between symbols (Bunge, 1962). A consensus of the scientific community is that the basic methods of science are *mathematical theory* and controlled *systematic experiment*. However, geographical systems especially human geographical systems are generally uncontrollable so that it is impossible to implement regular systematic experiments for geographical research. The experiment method is always replaced by experience of geographers. On the other hand, the application of traditional mathematical tools to geography is limited because of spatio-temporal complexity of geographical systems.

Mathematical theory is the first paradigm for scientific research. Mathematical methods have two basic functions in a discipline: one is to construct postulates and make models for developing theories, and the other is to possess and analyze experimental data or observational data. However, in quantitative revolution, most geographers preferred the second function to the first function, despite the fact that the first function is more important than the second function for development of geography (Moss, 1985). In fact, there are three obstacles to mathematical modeling for geographical systems, including spatial dimension, time lag, and interaction. All these difficult problems are related with scale dependence. The conventional mathematical tools such as calculous, linear algebra, probability theory and statistics can be applied to geographical analysis, but the conclusions are always inconsistent with the facts: the predicted values of models are often significantly different from the observed values, or the causalities cannot be efficiently revealed by the mathematical modeling. Some models, including the gravity model and allometric model, are useful in practice, but the model parameters cannot be interpreted with Euclidean geometry.

Because mathematical theory and systematic experiment cannot be efficiently applied to the traditional geographical research, geography once became of exceptional discipline that is different from the standard science (Schaefer, 1953). The biggest conundrum of geography development lies in *mathematization* of human geographical patterns and processes. Quantitative revolution resulted in quantification of human geography, which was regarded as becoming a real science (Stimson, 2008). Geography made great achievements in quantitative analysis and spatial modeling. However, the succeeding theorization of human geography is not really successful and quantitative human geography seems to turn a full circle (Johnston, 2008; Philo *et al*, 1998). The consequence of failure of geographical theorization is serious, which led to nihilistic mood in geographical circles. Atkinson and Martin (2000) once said, "Why, as geographers, would we want to throw away the geography?" (page 2) Earlier, Hurst (1985) once declared that "geography has neither existence nor future". In fact, the puzzles of geography boil down to two aspects. First, scale invariance. Traditional mathematical methods are based on characteristic scales, but geographical distributions are always free of characteristic scales; Second, irreducibility. Modern science is based on reductionism (Gallagher and Appenzeller, 1999; Waldrop, 1992), but geographical systems are actually irreducible. This paper is devoted to investigating the first aspect. In section 2, the key of mathematical modeling for geographical systems is clarified. In section 3, the new idea of geographical modeling based on scaling is illustrated. In section 4, the obstacles of geographical theorization and the solutions to these problems are discussed. Finally, the article is concluded with brief summary of this work.

## 2 The key of mathematical modeling

### 2.1 Characteristic scales

The key of key of quantitative measurement and mathematical modeling is to find characteristic scales, including characteristic length and characteristic parameters. Lord Kelvin once pointed out: "When you can measure what you are speaking about, and express it in numbers, you know something about it; but when you cannot measure it, when you cannot express it in numbers, your knowledge is of a meager and unsatisfactory kind." (see Taylor, 1977, page 37) However, not all measurements and numbers are valid for geographical analysis. If you want to measure a phenomenon or a system and express it in numbers, you must find the characteristic length, which represents the key scale of calculation or quantitative analysis. For example, for a circle, the characteristic length is its radius; for a square, the side length is its characteristic length; for a matrix, the eigenvalues give the characteristic lengths in different directions; for a statistical distribution, the average value is associated with the characteristic length. For a geometrical object such as circle, if we know the characteristic length, i.e., radius, we can know other geometrical information including the circumference and the area within the circle. For a statistical distribution, if we know the average value, we can compute the standard deviation and covariance, and the structure of probability become clear. A statistical inference is always made by means of probability structure comprising average value, standard deviation, and covariance.

Characteristic scales are basis of mathematical modeling and quantitative analysis by conventional methods. Generally speaking, a good mathematical model of a geographical system includes three levels associated with three scales: the basic level based on the characteristic scale, the macro level based on the global/large scale of geographical environment, and the micro level based on the local/small scale, which can be expressed with some of coefficients (Chen, 2008; Hao, 1986). An efficient and valid quantitative analysis is always based on the characteristic scale or the characteristic parameter of a model associated with the characteristic scale of the corresponding geographical system. For example, the rate parameter of Clark's model of urban population density is just the reciprocal of the characteristic radius (Chen, 2008a; Takayasu, 1990), and the Moran's index of spatial autocorrelation proved to be a characteristic root of a spatial correlation matrix based on the spatial weight matrix (Chen, 2013a). However, in many cases, we cannot find valid numbers indicating characteristic length for a geographical system, especially for a human geographical system. For instance, for the rank-size distribution of cities which follow Zipf's law, we can never find a valid mean to represent the characteristic length. Of course, we can calculate an average value, but the mean value depends on sampling results and sample sizes. We cannot find the fixed average value for the city-size distribution within a geographical region. Especially, the mean value does not represent the characteristic of the most elements in a sample. The calculated average value barely has no practical significance for us to explain the rank-size distribution of cities and predict urban development. This suggests an important concept: scale-free distributions. There is no typical size for a city in any country (Buchanan, 2000). It is meaningless to find the optimal city size for a regional system, but in theory, there is an optimal city size distribution (Chen, 2008a).

In short, there are two types of geographical phenomena, which correspond to two kinds of probability distributions. One is the *scaleful distribution*, which bears a characteristic scale; and the other is the *scale-free distribution*, which bears no characteristic scale (Table 1). The two sorts of distributions indicates two types of geographical space: scaleful geo-space and scale-free geo-space. The geographical phenomena satisfying a scaleful distribution can be treated with the conventional mathematical methods such as the higher mathematics. However, the geographical phenomena satisfying a scale-free distribution cannot be dealt with using the conventional mathematical tools (Figure 1). It needs the method of scaling analysis such as fractal geometry, allometric theory, and complex network theory (Batty, 2008; Batty, 2013; West, 2017).

**Table 1 A comparison between scaleful distribution example and scale-free distribution example**

| Order or step | Scaleful distribution: measurement result is independent of measuringscale | | Scale-free distribution: measurement result depends on measuringscale | |
|---|---|---|---|---|
| | Scale $x$ | Measurement $y_0$ | Scale $x$ | Measurement $f(x)$ |
| 1 | 1000 | 15 | 1000 | 9 |
| 2 | 100 | 15 | 100 | 16 |
| 3 | 10 | 15 | 10 | 28 |
| 4 | 1 | 15 | 1 | 50 |

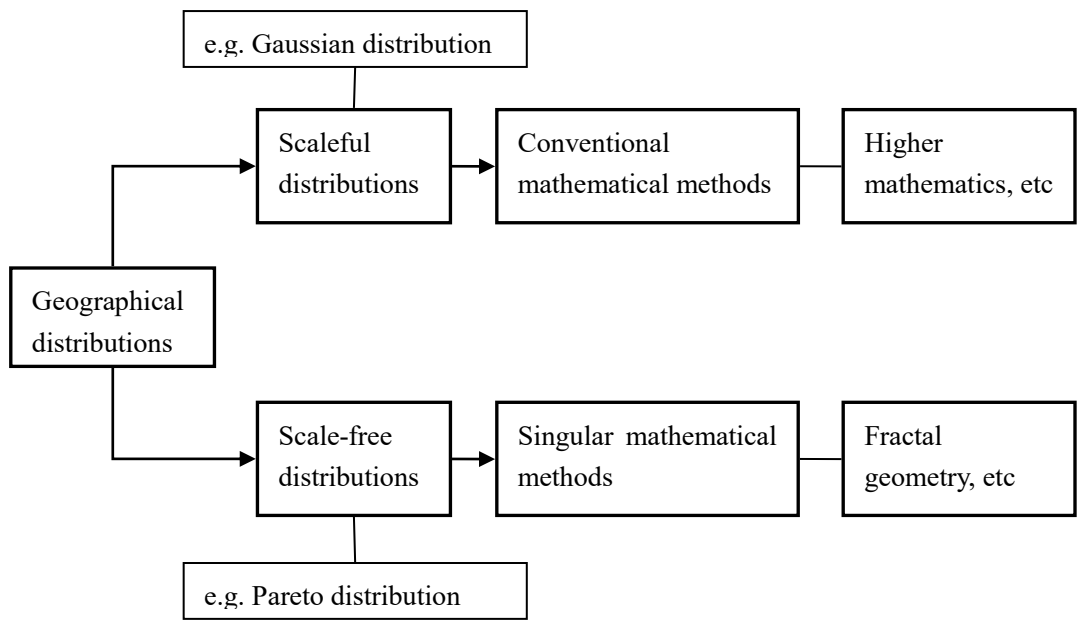

Figure 1. Two types of geographical phenomena corresponding to two types of probability distribution

## 2.2 Two types of spatial distributions

After quantitative revolution, geography evolved from a descriptive discipline into a science of spatial distributions. In the process of quantitative analysis, a spatial distribution of a geographical phenomenon can be converted into a statistical distribution using mathematical language and observational data. As stated above, statistical distributions fall into two types: *scaleful distributions* and *scale-free distribution* (Table 2). The typical scaleful distribution is normal distribution, which is often termed Gaussian distribution (Figure 2(a)). This is a very simple statistical distribution

$$f(x) = \frac{1}{\sqrt{2\pi}\sigma} e^{-\frac{(x-\mu)^2}{2\sigma^2}}, \qquad (1)$$

where $x$ refers to the scale and $f(x)$ to the probability density, $\mu$ denotes the average value of $x$, and $\sigma$ is standard deviation of $x$. The mean value $\mu$ indicates the characteristic length of the statistical distribution, and both the mean value $\mu$ and the standard deviation $\sigma$ define the probabilistic structure of the distribution. The representative geographical model is the normal function urban population density distribution (Dacey, 1970; Sherratt, 1960; Tanner, 1961). The typical scale-free distribution is the Pareto distribution (Figure 2(b)). This is a power-law distribution familiar to many scientists:

$$f(x) = Cx^{-b}, \qquad (2)$$

in which $C$ refers to the proportionality coefficient, and $b$ to the scaling exponent. For Pareto distribution, the average value of $x$ is invalid for directly quantitative analysis because it depends on the sample size. The urban traffic model of Smeed (1963) indicates an inverse power law distribution of road density around a city.

Table 2 A comparison between simple distributions and complex distributions

| Item | Scaleful distribution | Scale-free distribution |
|---|---|---|
| **Typical case** | Gaussian distribution | Pareto distribution |
| **Basic property** | With characteristic scale | Without characteristic scale |
| **Probability structure** | Can be defined with mean, standard deviation, and covariance | Cannot be defined with mean, standard deviation, and covariance |
| **Quantitative criterion** | Characteristic length | Scaling exponent |
| **Probability curve** | Unimodal curve | Long-tailed curve |
| **Geographical system** | Simple systems | Complex systems |
| **Example** | Urban population density distribution | City rank-size distribution |

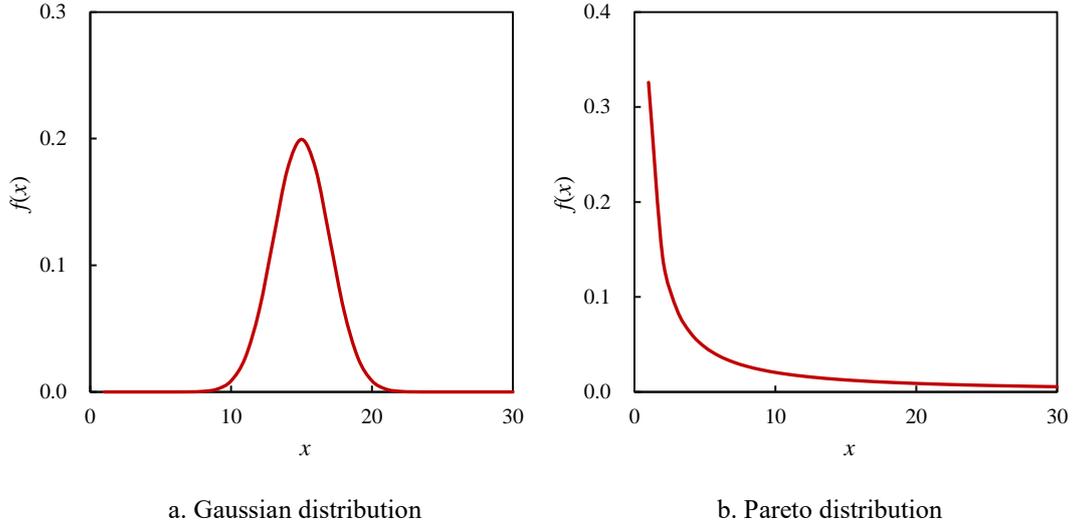

a. Gaussian distribution  b. Pareto distribution

**Figure 2. The distribution with characteristic scale and the distribution without characteristic scale**

(**Note**: The first one is the normal distribution with an average value of 15 and a standard deviation of 2. The second one is the power-law distribution with a scaling exponent of 1.2.)

If the probability density takes on a unimodal curve, it suggests a distribution with a characteristic scale. However, not all the scaleful distributions take on unimodal curves. For example, the exponential distribution takes on one-side decay curve with a thin tail rather than a single peak curve, but it is a scaleful distribution because the exponential distribution has a valid average value. The negative exponential function is as below:

$$f(x) = f_0 e^{-x/x_0},\qquad(3)$$

where $f_0$ refers to the constant coefficient, and $x_0$ to the scale parameter associated with a average value (Chen, 2008b). The well-known urban density model proposed by Clark (1951) is just a negative exponential function. Equation (3) is actually a density distribution function defined in 1-dimensional space. The curve based on equation (3) has no single peak. However, we can derive a unimodal curve defined in a 2-dimensional space from the 1-dimensional density distribution function (Chen, 2020). Compared with the power-law distribution, exponential distribution is simple (Barabasi and Bonabeau, 2003), while compared with the exponential distribution, the normal distribution is very simple (Goldenfeld and Kadanoff, 1999). In short, if a statistical distribution bears an average value indicating a characteristic scale, it is simple; while if a distribution has no valid mean indicative of characteristic scale, it is a complex distribution. Simple statistical distributions suggest simple geographical spatial or size distributions, while complex statistical distributions correspond to complex geographical spatial or size distributions. The simple distributions can be modeled with traditional mathematical methods. However, the complex distributions cannot be efficiently analyzed with higher mathematics.

## 3 New approach to mathematical modeling

### 3.1 Scaling and mathematical modeling

As indicated above, the key to quantitative measurement and mathematical modeling is the characteristic scale. If and only if we find the characteristic scale for a system through observational data, we can make an efficient mathematical model. For examples, for the normal distribution, equation (1), the characteristic scale is $x=\mu$; for the exponential distribution, equation (3), the characteristic scale is $x=x_0$. However, for a geographical system such as a city or a network of cities, we cannot find the characteristic scale in most cases. In other words, no characteristic length can be measured and expressed in numbers. For instance, for the power-law distribution, equation (2), no characteristic scale can found for $x$. This suggests that the traditional mathematical tools including calculus, linear algebra, probability theory and statistics cannot be directly and efficiently applied to modeling the systems without characteristic scales. In this instance, we need new notion of mathematical modeling for geographical systems and the *characteristic scale* should be replaced with *scaling*.

In scaling analysis, characteristic lengths are often replaced by scaling exponents. In order to show how to substitute scaling for characteristic scale, let's see a very simple example

$$A = \pi r^d, \qquad (4)$$

where $r$ is scale variable, $A$ is the corresponding measurement, $d$ is a power exponent. If $d=2$, we will have a formula for the area of a circle, $A=\pi r^2$, in which $\pi \approx 3.1416$. The radius $r$ can be determined and it is just the characteristic length, which can give the area and perimeter of the circle. On the other hand, the power exponent $d=2$ is known and shows no useful information for the geometrical object. However, for urban form, the situation is different, and equation (4) should be rewritten as

$$A = A_0 r^D, \qquad (5)$$

in which the coefficient $A_0$ is not necessarily equal to 3.1416, and $D$ is not equal to 2. In both theory and empirical work, we cannot find a certain radius for a city. Changing the radius $r$, we will have different urban area and perimeter. In other words, the urban measurements depend on the scale which is adopted. In this case, the radius cannot act as the characteristic length for the urban studies, but the power exponent can be used as a characteristic parameter, which is termed fractal dimension of urban form (Batty and Longley, 1994; Frankhauser, 1994; Frankhauser, 1998). Based on the 2-dimensional digital maps, the fractal dimension values come between 0 and 2. A low value ($D$<1) and high value ($D$→2) rarely appear, and most results are near 1.7 (Batty, 1991; Batty and Longley, 1994; Chen, 2010a). This suggest that a city's radius has no characteristic length (no valid fixed radius), but the scaling exponent based on the variable radius is constant and reflects spatial characteristics. The dual relationships between characteristic scale and scaling can be expressed as follows

$$A(r) = \pi r^d \rightarrow \begin{cases} A = \pi r^2, \ r \text{ is a characteristic length} \\ A(r) = A_0 r^D, \ D \text{ is a characteristic parameter} \end{cases}, \quad (6)$$

This shows the similarities and differences between Euclidean geometry and fractal geometry, each going it own way.

Nowadays, fractal geometry, allometric theory, and complex network theory have broken a new path to mathematical modeling of geographical systems. All these theories can be integrated into a new framework by the scaling concept (Batty, 2008; Chen, 2013b). We can compute the fractal parameters, or the scaling exponents, which represent new spatial characteristic quantities and reflect spatial characteristics of geographical systems from new perspective. In other words, a fractal has no characteristic length, but its fractional dimension has characteristic length. Analogously, a complex geographical system has no characteristic length, but its scaling exponents have characteristic length.

## 3.2 Fractal geometry for geographical spatial analysis

Among all the theory or mathematical methods based on scaling idea, fractal geometry is the most important one for geographical analysis. First, fractal geometry is a mathematical tool, which belongs to the first paradigm of scientific methods. The first paradigm is the most significant paradigm for scientific research. Second, fractal modeling can combine both numbers and patterns in the best way. It makes a good link between observational data and geographical patterns. Third, fractal measurement depends heavily on fractional dimension. The fractal dimension is just a spatial parameter and is powerful for geo-spatial analysis. A fractal includes three elements: form, chance, and dimension (Mandelbrot, 1977; Mandelbrot, 1982). These elements correspond to patterns, processes, and information of geographical systems (Table 3).

Table 3 The relationships between fractals and geographical research

| Fractal | Geographical systems | Geographical analysis |
|---|---|---|
| **Form** | Geographical patterns | Spatial distribution |
| **Chance** | Geographical process | Dynamical evolution |
| **Dimension** | Geographical information | Spatio-temporal analysis |

The famous difficult problems of mathematical modeling include spatial variables, time lag, and interaction. First, spatial variables always result in dimension puzzle of quantitative analysis. The well-known economist Arthur (1992) once said: "Only we don't usually consider the spatial dimension in economics much, so that makes economics a lot simpler." (see Waldrop, 1992, page 141) Unfortunately, geographers cannot evade and must face the problem of spatial dimension. Second, time lag always leads to nonlinearity. Where there is a time lag, there is a responding delay indicating nonlinear problem, which, generally speaking, cannot be solved using traditional mathematical methods (Chen, 2009). Third, interaction between multiple objects leads to complex dynamics. Interaction is one of the scientific difficult problems in the 20th century (Liu *et al*, 2003). All these conundrums are associated with scale dependence. Scale dependence suggests scale-free distributions, in which no characteristic length can be found directly by spatial measurements. Fractal geometry is one of the important and useful tools for solving the three problems abovementioned. That is, fractal geometry can be used to deal with dimensional relations, nonlinear processes, and scale-free distributions. In this sense, fractal geometry provides an efficient tool for geographical spatial modeling.

# 4 Geography: on the threshold of theoretical revolution?

## 4.1 The shortcomings of higher mathematics

In the conventional scientific research, the basic tools of mathematical modeling are the higher mathematics, including calculus, linear algebra, and probability theory and statistics. However, these mathematical methods cannot meet the needs of geographical research effectively. In fact, the higher mathematical methods are often incompatible with geographical analyses in nature (Table 4). First, calculus theory is based on regular geometry. However, geographical patterns need irregular geometry. Second, linear algebra is based on linear superposition principle. However, geographical process bears nonlinearity, which violates superposition property. Third, probability theory and statistics are based on the distributions with characteristic scales. However, geographical distributions are always free of scale. That is to say, geographical distributions are of scale invariance and follow scaling laws. Because of these insufficiencies of the traditional mathematical methods, the 'theoretical revolution' of geography after its 'quantitative revolution' is unsuccessful. In other words, geography has succeeded in quantification, but failed to achieve its goal of theorization (Philo *et al*, 1998).

**Table 4 The inconsistency between the higher mathematical methods and geographical analyses**

| Mathematics | Mathematical base | Geographical phenomenon |
|---|---|---|
| **Calculus** | Regular geometry | Irregular patterns |
| **Linear algebra** | Linear superposition principle | Nonlinear processes |
| **Probability theory and statistics** | Scaleful distributions | Scale-free distributions |

Many geographical phenomena cannot be described with the traditional higher mathematics. Applying a mathematical method based on characteristic scale to the geographical patterns and processes without characteristic scales always results in problems of oversimplification. This phenomenon indicates the *spherical chicken syndrome* (Lederman and Teresi, 1993; Kaye, 1989), or even what is called the "*'gene for' syndrome*" (Gallagher and Appenzeller, 1999). Recent years, a number of new mathematical tools have emerged, including fractal geometry, allometric scaling, complex network, renormalization group, wavelet analysis, artificial neural network, genetic algorithm, and chaotic dynamics. Among all these new mathematical methods based on the idea from scaling, fractal geometry is the most important and powerful tool for geographical analysis. The reasons are as below: First, fractal geometry can be employed to characterize the irregular patterns of geographical systems. Second, fractal geometry can be used to explore the nonlinear processes of geographical evolution. Third, fractal geometry can be adopted to describe the scale-free distributions of geographical phenomena. In short, fractal geometry is the most effective tool to explore nonlinearity, singularity and irregularity. This geometric tool can remedy the defects of the old higher mathematics where geographical research is concerned.

## 4.2 Reinterpreting traditional geographical models using new ideas

For geography, the achievements of quantitative revolution are very great. However, due to the confusion between the concept of characteristic scales and that of scaling, the value of many achievements has been buried for a long time. The essence of a theory rests with models, especially mathematical models (Holland, 1998). Geographers presented many important theories and made significant models during the period of quantitative revolution (e.g. see Haggett *et al*, 1977). Unfortunately, it used to be hard to develop further parts of these theories and models, some of which were even abandoned for a time not because they are not good in practice, but because they could not be interpreted with the notions of traditional mathematics. However, the inexplainable theories and models can be explained using the concepts from fractals and scaling. In a sense, a number of theoretical puzzles in human geography are dimensional conundrums, which cannot be solved with Euclidean geometry. The simple and typical examples are the law of allometric growth and the gravity model. The allometric scaling relation between urban area and population is as below:

$$A = aP^b = aP^{D_p / D_a}, \qquad (7)$$

where $A$ refers to urban area (the corresponding dimension is $D_a$), $P$ to city population size within this area (the corresponding dimension is $D_p$), $a$ denotes the proportionality coefficient, and $b=D_p/D_a$ is the allometric scaling exponent. The value of the scaling exponent is a well-known puzzle of traditional human geography. The dimension of urban area used to be regarded as $D_a=d=2$. According to the principle of dimension consistency based on the idea of Euclidean geometry, if the dimension of city population is $D_p=3$, then $b=2/3$; if $D_p=2$ as given, then $b=1$ (Lee, 1989; Longley *et al*, 1991; Nordbeck, 1971). In short, the predicted value of $b$ by the traditional theory is either 2/3 or 1. However, the observed value of $b$ always come between 2/3 and 1 rather than either 2/3 or 1 (Chen, 2008a). The conflict between the theoretical value and the calculated value cannot be interpreted by Euclidean geometry, but it can be readily construed by fractal geometry (Batty and Longley, 1994; Longley *et al*, 1991; Chen, 2010b; Chen and Xu, 1999). The gravity model is familiar to geographers and the basic and classical form is

$$I_{ij} = G \frac{P_i P_j}{r_{ij}^{\alpha}}, \tag{8}$$

where $I_{ij}$ denotes the gravitation between places $i$ and $j$, $r_{ij}$ is the distance between the two places, $P_i$ and $P_j$ indicates the "masses" (size measurements) of places $i$ and $j$, respectively, $G$ refers to the gravity coefficient, and $\alpha$ to the distance-decay exponent. The exponent $\alpha$ cannot be interpreted with the dimensional concept based on Euclidean geometry so that the power-law impedance function was replaced by the exponential impedance function (Haggett *et al*, 1977; Haynes, 1975). However, the spatial interaction model based on the exponential function brought on a new problem, that is, the gravity model is based on the concept of action at a distance, but the exponential function suggests an effect of spatial localization instead of action at a distance (Chen, 2008b). This problem can also be solved by the concepts of fractal dimension (Chen, 2009; Chen, 2015).

The ideas from fractals and scaling can be used to reinterpret and develop many classical theories and models. These years, a number of important geographical theories and models have been improved or reinterpreted by using the concepts from fractals, including central place theory (Arlinghaus, 1985; Arlinghaus and Arlinghaus, 1989; Batty and Longley, 1994; Chen, 2011; Chen, 2014; Chen and Zhou, 2006), spatial interaction models (Chen, 2015), and spatial autocorrelation analysis (Chen, 2013a). Twenty year ago, Batty (1992) once observed: "Many of our theories in physical and human geography are being reinterpreted using ideas from fractals and tomorrow they will become as much a part of our education and experience as maps and statistics are today." (page 36) The above prediction has been being supported or even confirmed over and over again.

## 4.3 New tools for geographical research

Scientific research is nothing more than two processes: description and understanding. First, describe the characteristics of a system, and then try to understand its work mechanism (Gordon, 2005). In order to describe a phenomenon precisely, we need mathematics and measurements, and in order to understand the mechanism deeply, we need laboratory experiments and computer simulation (Henry, 2002; Waldrop, 1992). Effective geographic spatio-temporal description and understanding can lead to effective interpretation and prediction. As Fotheringham and O'Kelly (1989, page 2) once pointed out: "all mathematical modelling can have two major, sometimes contradictory, aims: explanation and prediction." In fact, the main functions of science lie in explanation and prediction (Kac, 1969).

The traditional paradigms of scientific research include mathematical theory (originated in ancient Greece) and laboratory experiment (originated in the Renaissance). The third important paradigm is computer simulation originated after World War II (Bak, 1996). Before the turn of the century, "there were three ways now to proceed in science: mathematical theory, laboratory experiment, and computer modeling" (Waldrop, 1992, page 268). Today, scientists are telling us the fourth paradigm has been emerging, and the paradigm can be termed data-intensive computing (Bell *et al*, 2009; Hey *et al*, 2009). Today, scientific paradigms have developed, including mathematical theory, experience and (laboratory) experiment, computer simulation, and data-intensive computing. In each paradigm of scientific methodology, we can find new tools for geographical research and spatial analysis (Table 5). Of course, laboratory experiments are still an exception. However, we can carry out on-the-spot investigation in light of the new way of geographical thinking so as to make up for the lack of geographical research which cannot be made by laboratory experiment.

**Table 5 The four paradigms in science and their applications to geographical research**

| Type | Paradigm | Function | Geographical research | New tools for geography |
|---|---|---|---|---|
| **The first paradigm** | Mathematical theory | Data processing and theoretical modeling | Processing observational data | fractal geometry, nonlinear mathematical methods, etc. |
| **The second paradigm** | Laboratory experiment and experience | Finding out causal relationships | Geographical systems are uncontrollable and the laboratory experiment is replaced by experience | field investigation based on new thinking from fractals and scaling |
| **The third paradigm** | Computer simulation | Finding out causal relationships and computer aided modeling | It can be used to make up for the deficiency of laboratory experiment | cellular automata (CA), multiple agent system (MAS), etc. |
| **The four** | Data- | Big data | Processing big data | scaling analysis, |

| **paradigm** | intensive computing | processing | | including allometry and complex network |

## 5 Conclusions

Due to the shortage of methodology, the theoretical development of geography lags behind for a long time. However, things will change in the near future. Now, the barriers of theorization of geography are being partially removed, and to my thinking, geography may be on the threshold of theoretical revolution. The main viewpoints of this work can be summarized as follows. **First, fractal geometry and a number of mathematical methods based on the ideas of scaling provide new tools for geographical description.** A great many geographical phenomena have no characteristic scales, and cannot be characterized by means of conventional mathematical methods. Replacing characteristic length with scaling exponents, we can make spatial analyses of scale-free geographical systems from a new perspective. In particular, fractal geometry is the most important and powerful mathematical tool for geographical analysis among various new mathematical methods. The main difficulties of geographical mathematical modeling include spatial dimension, time lag, and interaction. Fractal geometry is one of the significant and useful tools for solving these problems. Effective description leads to effective understanding, which in turn results effective explanation and prediction. **Second, computer modeling and simulation can be used to explore the causalities in geographical spatio-temporal processes.** The geographic systems are uncontrollable, so we can't analyze its evolution mechanisms with the help of systematic controlled experiments. Using computer simulation, we can find out the causal relationships hidden behind geographical spatial patterns and evolution processes. Computer simulation must be closely combined with mathematical modeling. Without valid theoretical modeling, computer simulation is blind, while without computer simulation, mathematical models may become lame. The traditional thinking of geographical mathematical modeling is based on the characteristic scales, and the corresponding computer simulation is also based on characteristic scale notion. In future, the idea of scaling should be introduced into computer simulation of geographical evolution. **Third, the ideas and theories based on scaling can be employed to improve old theory and develop new theory in geography.** As indicated above, many theories in physical and human geography can be reinterpreted and improved using ideas from fractals and scaling. Especially, new models can be built and new theories can be developed by integrating ideas based on characteristic scales and those based on scaling. Complex geographic systems have two sides of unity of opposites: some phenomena bear characteristic scales, and some phenomena have no characteristic scales. In order to understand different geographical processes, we must adopt proper mathematical methods to describe different geographical phenomena. Traditional mathematical tools are not suitable for scaleful geographical phenomena, just as scaling analysis cannot be used for scale-free geographical phenomena. Only when the methods are used properly can good results be achieved in geographical modeling and analyses.

**Acknowledgements**

This research was sponsored by the National Natural Science Foundations of China (Grant No. 41671167). The support is gratefully acknowledged.